# Docking Peptides into HIV/FIV Protease with Deep Learning and Focused Peptide Docking Methods

Katherine Ge[1,2], Dayna Olson[1,3], and Michel F. Sanner[1]

[1] *Scripps Research, 10550 N Torrey Pines Rd, La Jolla, CA 92130, USA*
[2] *The Bishop's School, 7607 La Jolla Blvd, La Jolla, CA 92037, USA*
[3] *Michigan State University, 220 Trowbridge Rd East Lansing, MI 48824, USA*

## Abstract

Molecular docking is a structure-based computational drug design technique for predicting the interaction between a small molecule (ligand) and a macromolecule (receptor). Over the past three decades various docking software programs have been developed, mostly for drug-like molecules. With the recent interest in peptides as therapeutic molecules, several peptide docking methods have also been developed. AutoDock CrankPep (ADCP), is a state-of-the-art peptide docking tool that predicts the interaction of peptide with up to 20 amino acids in a user defined region of a macromolecule, i.e.focused docking. Recent advances in deep learning (DL) approaches have shown remarkable success in docking linear peptides composed of natural amino acids only. Unlike ADCP, these methods provide a confidence level in their predictions. Here we explore whether ADCP and various DL methods (AlphaFold2 Monomer, AlphaFold2 Multimer, and OmegaFold) and their prediction confidence metric can be used to discriminate native and non-native substrates for HIV and FIV proteases. We found that ADCP successfully predicts the interactions of native peptides but fails to discriminate non-native ones. This was expected as conventional docking methods report solutions maximizing ligand receptor interactions for any ligand. Surprisingly, DL methods underperform when docking native peptides into these particular docking targets but achieve high success rates with non-native peptides. While AlphaFold managed to successfully dock a few of the native peptides, OmegaFold failed to successfully dock any of them. Overall, none of these methods is currently able to distinguish between native and non-native peptides, warranting further exploration of specialized methodologies.

## Introduction

Therapeutic peptides are a category of pharmaceuticals composed of short amino acid chains. They provide highly targeted and precise interactions with biological targets. While earlier peptides had limitations related to stability and bioavailability, recent advancements in peptide design and delivery technologies have significantly improved their therapeutic potential[1-11].

Peptide docking refers to the computational process used to predict the complex interactions between peptides and proteins. Although accurate prediction is complex due to the size, flexibility, and intricate nature of peptides[12][13], recent enhancements in algorithms and computational power have led to significant improvements in peptide docking methods. These methods can be categorized into focused docking, used when the binding site is known, and blind docking, used when the binding site is unknown [14-17]. A state-of-the-art focused peptide docking method[6] is AutoDock CrankPep (ADCP) [18][19] utilizes a rigid receptor and flexible ligand to optimize the interaction between the two, and offers state-of-the-art accuracy for conventional focused peptide docking methods [20][21]. Focused peptide docking methods often identify accurate poses within their top 10 to 100 solutions, yet they reach an approximate success rate of only 20% when considering solely the top-ranked docked poses. Recently, deep learning (DL) methods have begun to be widely applied to blind peptide docking. AlphaFold2 [22][23][24], initially designed for predicting monomeric protein structures, was recently demonstrated by Tsaban et al. to be surprisingly effective at predicting protein-peptide complexes[25][26]. This was achieved by using AlphaFold2 to fold a single chain comprising the protein and peptide sequences linked by a lengthy polyglycine linker - an approach we'll refer to as AF2mono. With a non-redundant set of 96 complexes, this method predicted half the complexes within a 2.5Å peptide backbone RMSD. A new iteration of AlphaFold2 models, referred to as AlphaFold2 multimer [27] (AF2multi), has been released. These models, while retaining the same deep neural network architecture as their AF2mono predecessors, have been specifically trained to predict the structures of multimeric proteins. OmegaFold[28] is another DL peptide docking method that folds proteins based on amino acid sequence and a geometry-inspired protein language model without explicitly building sequence alignment making it fast.

A recent study compared DL-based methods and AutoDock CrankPep for predicting protein-peptide interactions using a dataset of 99 protein-peptide complexes[29]. The study found that AF2multi significantly outperformed AF2mono, OmegaFold, and ADCP. Interestingly, ADCP and AF2multi showed complementary abilities in terms of the types of peptides they could successfully dock. The DL methods were able to predict instances where certain peptides might not interact with the target protein, a capability not always present in conventional methods. However, in scenarios where the peptide's binding site is known, DL methods currently cannot leverage this information. It remains intriguing to compare DL methods to a focused docking approach in such cases. Additionally, the study introduced new metrics that may aid in exploring a docking method's ability to differentiate between binders and non-binders, addressing a common issue in conventional docking methods that always report the best identified binding mode.

In this study, our objective is to evaluate and compare the capability of DL blind peptide-docking methods (AF2mono, AF2multi, OmegaFold ) and conventional focused peptide-docking method ADCP in distinguishing between native and non-native substrates for HIV and FIV proteases. We chose these proteases as they are known to cleave the HIV polyprotein in multiple sequences and are well represented in the datasets on which the DL methods were trained. We designed non-native sequences by selecting amino acids in each position reversing the chemico-physical properties observed for amino acids in this position in the natively cleaved peptides in order to maximize the probability of this sequence not to fit the protease binding site, thus avoid cleavage of this sequence. We then docked the native, and designed sequences into HIV and FIV protease using OmegaFold, AlphaFold monomer AlphaFold multimer, and

AutoDock CrankPep, and assess the ability of these software programs to properly predict the protein peptides interactions. For native sequences we expect the cleaved peptide bond to be within the correct distance of the protase's catalytic residues. For non native sequences, we expect that none of the peptide's peptide bonds is close enough from the catalytic residues to be cleaved.

## Methods

### Dataset

For HIV protease, we choose 10 native peptides [30] cleaved in Gag-Pol polyprotein and 9 designed sequences (non-native peptides) [31]. For FIV, we choose 7 native peptides [30] and 7 designed sequences. The design of synthetic sequences involved analyzing the physicochemical properties (such as polarity, size, charge, pH sensitivity, etc.) of native peptides and selecting amino acids with opposite physicochemical attributes to minimize chances of being cleaved.

Table 1: Peptide Sequences for HIV Protease

| Site | Native | Site | Designed |
|---|---|---|---|
| MACA | VSQNYPIVQN | seq1 | WDVDQTHMER |
| CAP2 | KARVLAEAMS | seq2 | GEAEHSRKDH |
| P1NC | PATIMMQRGN | seq3 | GDAKREKQED |
| P1P6 | RPGNFLQSRP | seq4 | HEVRKQKRDE |
| INP6 | DKELYPLTSL | seq5 | FDVHECHMEK |
| TFPR | VSFNFPQITL | seq6 | NEVMDKRWER |
| PRRT | CTLNFPISPI | seq7 | GDAQDRHHDR |
| INRT | GAETFYVDGA | seq8 | NEAQENRWDH |
| RTIN | IRKILFLDGI | seq9 | WDVEKDKHEK |
| NCP2 | ERQANFLGKI | | |

Table 2: Peptide Sequences for FIV Protease

| MACA | PPQAYPIQTV | seq1 | EEIRDEYWKR |
|---|---|---|---|
| CANC | LTKVQVVQSK | seq2 | YFECDSQDME |
| NCPR | IGFVNYNKVG | seq3 | CSDTIPGEDQ |
| PRRT | IRLVMAQISD | seq4 | DECYQNWDVK |
| RTRN | GAETWYIDGG | seq5 | STKLNMYRQC |
| RNDU | LCQTMMIIEG | seq6 | YAFPKKSTRN |
| DUIN | STGVFSSWVD | seq7 | LAKTRVGRNK |

Letter colors defined in peptide sequences table : Cleavage site (red), 7-mer (black, red), 10-mer (blue, black, red)

**Peptide Docking Methods**

*AlphaFold2 Monomer*: Docking was carried out by joining the macromolecule–either the HIV or FIV protease–to the peptide sequence using a 30-residue-long glycine linker. AF2 Mono was run using default settings, generating 5 complexes, one per AF2 model. In postdocking processing, the polyglycine linkers were removed and the monomeric poses split into two chains.

*AlphaFold2 Multimer:* Docking was carried out using 2 separate sequences for the macromolecule and the peptide. AF2 Multi also yields 5 complexes.

*OmegaFold:* This method uses a single sequence as input but does not compute a multiple sequence alignment (MSA). Like the AF2 Monomer, peptide and receptor sequences are concatenated with a 30 glycine linker. OF generates 1 complex per input sequence. Also like AF2 Monomer, postdocking processing required the glycine linker be removed and the monomeric sequence split into two.

*AutoDock CrankPep:* AutoDock affinity maps were generated using AutoGridFR (AGFR) with default parameters and a docking box centered on the peptide with 7Å padding in every direction. AGFR produced a target file (.trg) that is used as the receptor when docking with ADCP. Docking was performed with default settings for ADCP (3 million evaluations of the scoring function per amino acid in the peptide, for each Monte Carlo search). ADCP used protease structure 2i4w.pdb (HIV) and 2hah.pdb (FIV) from the Protein DataBase. It generated 100 solutions per complex.

**Docking Accuracy Assessment Method**
Successful metric is defined to evaluate docking success. For native peptides, we claim a success if Oxygen atoms in the catalytic ASP25 side chains are within 5Å of cleaved peptide bonds. For non-native peptides we claim a sucess if there is no peptide-bond within 5Å of the oxygen atoms in the catalytic ASP25 side chains. We compare both the success rate and simulation time of those algorithms.

## Results and Discussion

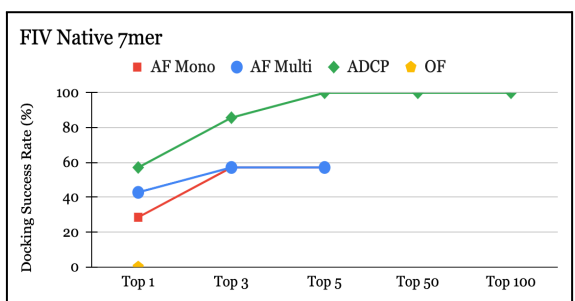
(a) FIV native peptide of 7 mer

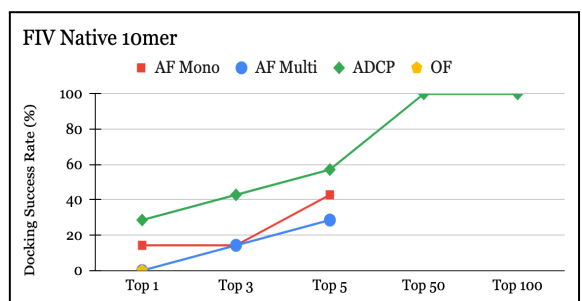
(b) FIV native peptide of 10 mer

Figure 1: Compare prediction accuracy for FIV native peptide

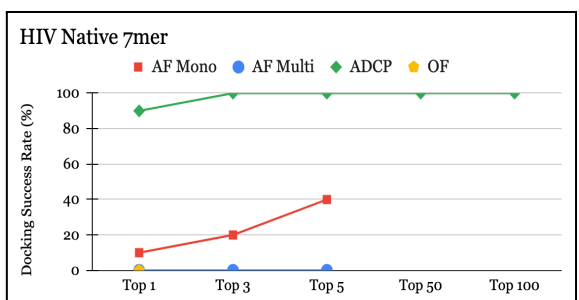
(a) HIV native peptide of 7 mer

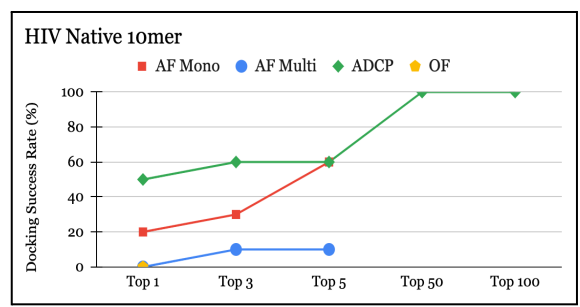
(b) HIV native peptide of 10 mer

Figure 2: Compare prediction accuracy for HIV native peptide

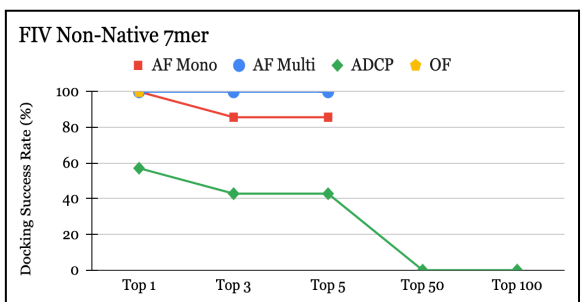
(a) FIV non-native peptide of 7 mer

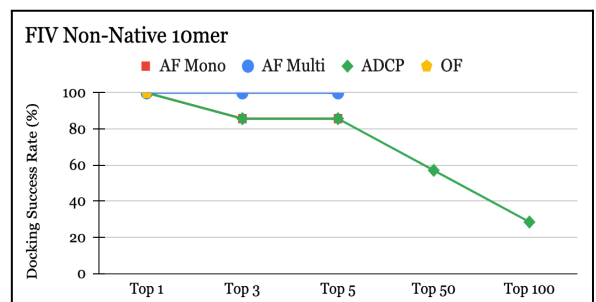
(b) FIV non-native peptide of 10 mer

Figure 3: Compare prediction accuracy for FIV non-native peptide

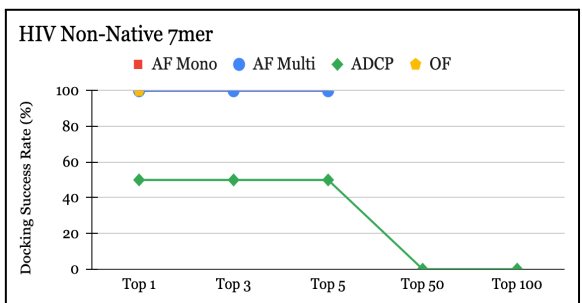 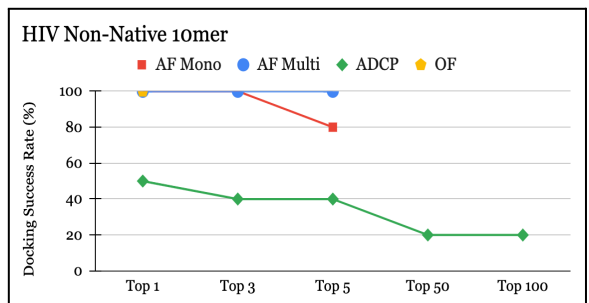

(a) HIV non-native peptide of 7 mer　　　　　　(b) HIV non-native peptide of 10 mer

Figure 4: Compare prediction accuracy for HIV non-native peptide

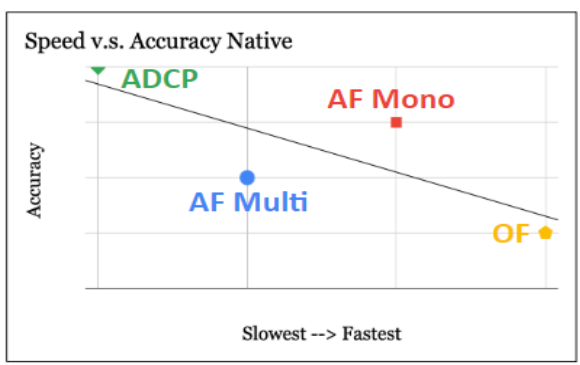 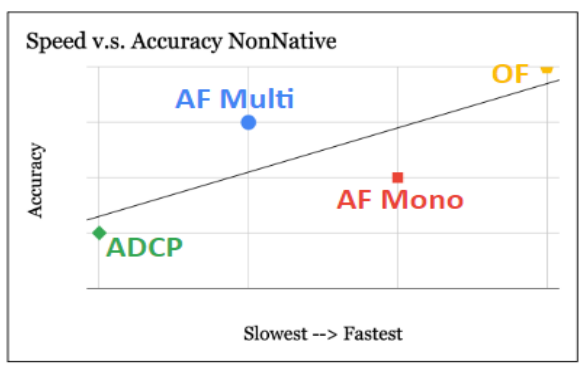

(a) Native peptide　　　　　　(b) Non-native peptide

Figure 5: Compare speed vs accuracy among ADCP, AF Mono, AF multi and OF

For native peptide docking, as shown in Figure 1 and 2, our findings reveal that ADCP, a focused peptide docking tool, exhibits notably superior performance compared to deep learning (DL) methods. Additionally, contrary to expectations, AF Mono outperforms the AF multi in this context. Strikingly, OmegaFold failed to successfully dock any of the native peptides accurately. These unexpected results shed light on the distinct strengths and limitations of different docking approaches in the context of native peptide substrates.

Regarding non-native peptides, as shown in Figure 3 and 4, the DL methods exhibit remarkable performance, which may be attributed to the possibility that these methods are inherently better suited for docking peptides into the studied proteases. As anticipated, ADCP demonstrates suboptimal performance with non-native sequences due to its tendency to prioritize solutions that maximize interactions within the protease tunnel. This aligns with our expectations since ADCP was specifically designed for focused docking within the protease tunnel and may struggle when confronted with non-native peptide sequences.

Although OmegaFold (OF) is the most computationally efficient approach, as it only produces one complex, it fails to generate any poses within 5Å for both native and non-native. Our findings suggest that it may not be suitable for accurately predicting peptide-protein interactions. This highlights the importance of considering not only computational cost but also the predictive capabilities and accuracy of a docking approach when selecting the most appropriate method for peptide-protein interaction prediction.

The AF2 models exhibit subpar performance than ADCP when applied to these proteases. However, the deep learning (DL) methods impressively fold the receptor accurately, even for FIV, which is comparatively less represented in the DL training set. This is evident from the backbone root-mean-square deviation (RMSD) values, which are consistently below 1.8Å. These results underscore the ability of DL methods to effectively predict the receptor conformation, highlighting their potential in peptide docking studies, despite their underperformance in the specific context of these proteases.

In contrast to the more challenging task of blind docking, where a peptide is docked into a receptor of unknown structure, ADCP tackles a relatively simpler problem by docking into an already folded receptor and focusing on the known binding site. ADCP demonstrates impressive accuracy in predicting the docked poses of native peptides. This highlights the robustness and efficacy of ADCP in capturing the intricate interactions between peptides and their target receptors, particularly in cases where the receptor structure is known.

## Conclusion

The ability of physics-based methods to to dock peptides currently out of reach for Deep Learning (DL) peptide docking methods, such as cyclic peptides or those containing non-standard amino acids, or being covalently bound, makes these methods highly complementary. We also previously demonstrated the synergistic complementarity of AutoDock CrankPep and AlphaFold2 multimer in the case where both methods can be applied. This case study further underscores this complementarity between these methods and revealed that: i) DL peptide docking methods perform surprisingly poorly when predicting interactions between various peptides and HIV protease and FIV protease. While our physics-based method AutoDock CrankPep performed very well for native peptides, both methods failed at distinguishing between native and non-native peptides. This distinction remains a significant challenge in the field of peptide docking and requires further exploration and development of specialized methodologies.

## Acknowledgment & Bibliographies

The research reported in this publication was supported by the National Institute of General Medical Sciences of the National Institutes of Health under Award Number R01GM096888 to Dr. Michel F. Sanner. This is manuscript 30249 from The Scripps Research Institute.